\documentclass[prb,aps,twocolumn,superscriptaddress,showpacs]{revtex4}
\usepackage{epsfig}
\usepackage{amsmath}

\begin{document}

\author{E.~V.~Bezuglyi}
\affiliation{B. Verkin Institute for Low Temperature Physics and
Engineering, Kharkov, 61103 Ukraine}

\author{V.~S.~Shumeiko}
\author{G.~Wendin}
\affiliation{Chalmers University of Technology
and G\"oteborg University, S-41296 G\"oteborg, Sweden}

\title{ Nonequilibrium Josephson effect in short-arm diffusive SNS
interferometers. }

\date{\today}

\begin{abstract}

We study non-equilibrium Josephson effect and phase-dependent
conductance in three-terminal diffusive interferometers with short
arms. We consider strong proximity effect and investigate an
interplay of dissipative and Josephson currents co-existing within
the same proximity region. In junctions with transparent interfaces,
the suppression of the Josephson current appears at rather large
voltage, $eV\sim \Delta$, and the current vanishes at $eV\geq\Delta$.
Josephson current inversion becomes possible in junctions with
resistive interfaces, where it occurs within a finite interval of the
applied voltage. Due to the presence of considerably large and
phase-dependent injection current, the critical current measured in a
current biased junction does not coincide with the maximum Josephson
current, and remains finite when the true Josephson current is
suppressed. The voltage dependence of the conductance shows two
pronounced peaks, at the bulk gap energy, and at the proximity gap
energy; the phase oscillation of the conductance exhibits
qualitatively different form at small voltage $eV<\Delta$, and at
large voltage $eV>\Delta$.

\end{abstract}
\maketitle

\section{Introduction}

\begin{figure}[tb]
\epsfxsize=8.5cm\epsffile{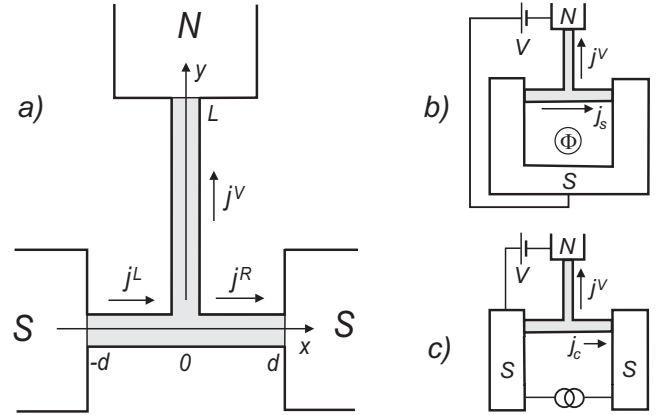}
\caption{Sketch of 3-terminal SNS interferometer: T-shape diffusive
metallic wire with short arms (shaded region) is connected to
superconducting (S) and normal (N) reservoirs (a), flux bias setup
for measuring current-phase relation (b), current bias setup for
measuring critical current (c). }
\label{device} \vspace{-4mm}
\end{figure}

Multiterminal superconductor - normal metal - superconductor (SNS)
junctions are interesting devices where an interplay between the
dissipative normal electron current and non-dissipative Josephson
current can be studied. The simplest device of this type consists of
two superconducting reservoirs and one normal reservoir connected by
a small normally conducting T-shape bridge, see Fig.~\ref{device}(a).
A mesoscopic size of the bridge is essential to keep the coherence of
the current transport over the whole device. During the last decade,
a large amount of interesting experiments have been done with such
kind of devices (for the review see Ref.~\onlinecite{Lambert} and
references therein, further references can be found in
Ref.~\onlinecite{Sam}).

Non-equilibrium state in multiterminal SNS junctions exhibits two
closely related major phenomena: the interferometer effect, which
concerns the dependence of normal conductance of the device on the
phase difference between the superconducting
reservoirs,\cite{Hekking} and the Josephson transistor effect, which
concerns the dependence of the Josephson current on the current
injected from the normal reservoir.\cite{Wees,Wendin} The
interferometer effect gives rise to a number of so called
Josephson-like effects.\cite{VP,VT}

The interferometer effect has received the most of attentions; it has
been extensively studied experimental\-ly\cite{Petrashov,Lambert} and
theoretically,\cite{VZ,Lambert,Sam} and presently this effect is
rather well understood. The proximity of the superconducting
reservoirs leads to a modification of the density of states and
transport properties of the normal bridge (proximity effect), which
therefore become sensitive to the phase difference at the reservoirs,
and exhibit oscillating behavior as the function of the phase
difference.

The Josephson transistor effect has been also observed in
experiment.\cite{Morpurgo98,Baselmans02,Huang} The non-equilibrium
Josephson effect has been first predicted\cite{Wees} and then
theoretically studied in the ballistic
junctions.\cite{SamRap,Bagwell,SamPhysC,Sam} Non-equilibrium
population of the Andreev states\cite{Andreev} induced by the current
injection leads to a full-scale variation of the Josephson current
with the applied voltage and to the inversion of the current
direction. Similar transistor effect has been also investigated in
diffusive junctions.\cite{VolkovSINIS,Wilhelm98,Yip,Heik}

The most of experiments with multiterminal junctions have been done
using diffusive metallic bridges, whose length was large compared to
the superconducting coherence length $\xi_0$. In such long junctions,
the proximity effect is suppressed, which results in small amplitude
of the conductance oscillation (typically few percents of the full
conductance value), and small magnitude of the Josephson current. In
theoretical analysis of the interferometer effect in diffusive
junctions, the presence of the Josephson current is usually ignored.
Similarly, the theoretical studies of the nonequilibrium Josephson
effect\cite{VolkovSINIS,Wilhelm98,Yip} are restricted to the regime
of ``weak'' proximity effect,\cite{VZ,VK,VolkovI} when the induced
gap in the normal bridge is much smaller than those in the
superconducting reservoirs. Such a regime is relevant for long
diffusive junctions and for junctions with high-resistance NS
interfaces. In addition, the non-equilibrium Josephson current was
calculated for specific four-terminal circuit geometries where the
dissipative normal current and the Josephson current were spatially
separated.

Meanwhile, it is conceptually interesting to investigate the problem
of interplay of dissipative and Josephson currents flowing in the
same diffusive lead under a strong proximity effect. In this paper,
we address this problem by studying the non-equilibrium Josephson
effect in 3-terminal SNS interferometers with short arms, whose
lengths are {\em smaller} than the coherence length $\xi_0$.

Co-existence of dissipative and non-dissipative currents in the same
proximity region makes it difficult to identify the Josephson current
component. In equilibrium, the total current flowing through a
proximity lead is the Josephson current, which is entirely determined
by the supercurrent spectral density and the population numbers of
the relevant states. The presence of the normal injection current and
related gradients of the distribution functions violates the local
conservation of the supercurrent component which varies along the
lead, and hence its direct connection to the Josephson current is
lost. Nevertheless, as we will show in the paper, a simple picture of
the non-equilibrium Josephson effect as the result of non-equilibrium
population of the states with the same current spectral density as in
equilibrium can be justified for some particular cases (see also
Ref.~\onlinecite{Huang}). However, a general situation seems to be
more complex.

In the short-arm SNS interferometers, the proximity effect is strong
when the interfaces are transparent, but it also may be strong when
the interfaces are highly resistive. The measure of the strength of
the proximity effect is the magnitude of the induced energy gap,
which is comparable with the superconducting energy gap in the
reservoirs. Consequently, there are full-scale variations of the
Josephson current and the normal conductance with the applied voltage
and the phase difference.

In short diffusive junctions with transparent interfaces, the
Josephson current is solely carried by the Andreev states whose
energies are smaller than the superconducting energy gap, which is
similar to the short ballistic junctions.\cite{Fur,Wendin}
Consequently, in these junctions, the Josephson current can be
suppressed to zero but never be reversed. The Josephson current
reversion becomes possible in the junctions with resistive interfaces
due to the negative contribution of the states with energies above
the bulk energy gap. Moreover, in contrast to weak proximity
regime,\cite{VolkovSINIS} the current reversion exhibits a fine
structure similar to the one theoretically discussed for ballistic
junctions,\cite{SamPhysC} and recently demonstrated in the experiment
with long diffusive junctions.\cite{Baselmans02}

The interplay of the injected and Josephson currents in the strong
proximity regime is important for the interpretation of experiments
with current biased junctions, Fig.~\ref{device}(c). The magnitude of
the non-equilibrium critical Josephson current measured with this
setup is different, as we will show, from the one derived from the
current-phase relation measured by an rf-SQUID, Fig.~\ref{device}(b).
This phenomenon has been earlier noticed in the absence of true
Josephson current in the weak proximity regime.\cite{VP}

The structure of the paper is the following. After introducing a
basic formalism in Section II, we discuss the spectral functions in
Section III, and the distribution functions in Section IV. Sections V
and VI are devoted to a discussion of the non-equilibrium Josephson
effect; the interferometer effect is considered in Section VII.


\section{Basic equations}

The junction we are going to investigate is sketched in
Fig.~\ref{device}(a). It consists of two superconducting reservoirs
and a normal reservoir connected by mesoscopic T-shape diffusive
metallic bridge. Such a geometry can be realized in experiment, e.g.,
by using nanowire technology.\cite{nano} The superconducting
reservoirs are assumed to have equal potentials, and the
superconducting phase difference between the reservoirs is $\phi$.
The distance between the superconductors, $2d$ ($-d<x<d$), and the
length of the injection lead, $L$ ($0<y<L$), are assumed to be small
compared to the superconducting coherence length $\xi_0 =
\sqrt{\hbar{\mathcal D}/\Delta}$ ($\mathcal{D}$ is the diffusion
coefficient), however, the relation between these lengths can be
arbitrary. For simplicity, we assume the cross sections and normal
conductivities of all wires to be equal, and the current from the
voltage-biased normal reservoir to be injected in the middle of the
SNS junction.

Neglecting spatial variations of all quantities across the leads, we
use one-dimensional static equations\cite{LO} for the $4\times 4$
matrix Keldysh-Green function $\check{G}$ in the normal leads, in
which we neglect the inelastic collision term,
\begin{eqnarray}
&\displaystyle \left[\sigma_z E,\check{G}\right]=i \hbar{\mathcal
D}\partial \check{J}, \quad \check{J} = \check{G} \partial \check{G},
\quad {\check{G}}^2=\check{1}, \label{Keldysh}
\\
&\displaystyle \check{G}=\left(\begin{array}{cc}
\hat{g}^R & \hat{G}^K \\
0 & \hat{g}^A
\end{array}\right),
\quad \hat{G}^K = \hat{g}^R \hat{f} - \hat{f}\hat{g}^A.
\label{Keldysh1}
\end{eqnarray}
Here $\hat{g}^{R,A}$ are the retarded and advanced Green functions,
$\hat{f} = f_+ + \sigma_z f_-$ is the matrix distribution function,
and $\partial$ denotes spatial derivative. At the junction node, the
matrix current $\check{J}$ obeys the Kirchhoff's rule, \cite{Naz}
\begin{equation} \label{BoundaryJ}
\check{J}_{x=-0} =  \check{J}_{x=+0} + \check{J}_{y=+0}.
\end{equation}
The Keldysh component $\hat{J}^K$ of the matrix current $\check{J}$
determines the electric currents in the leads,
\begin{equation} \label{current}
j= {\sigma \over 4e}\int_0^\infty dE \mathop{\textrm {Tr}}\sigma_z
\hat{J}^K = {\sigma \over e}\int_0^\infty dE \,I_-(E),
\end{equation}
where $\sigma $ is the normal conductivity. The current spectral
density $I_-(E)$ in Eq.~(\ref{current}) has three
components,\cite{VolkovI}
\begin{equation} \label{currents}
I_- \equiv (1 / 4)\mathop{\textrm {Tr}}\sigma_z\hat{J}^K   =
D_-\partial f_- + I_s f_+ - I_{\mathit{an}} \partial f_+,
\end{equation}
The first term in Eq.~(\ref{currents}) describes a dissipative
current which provides usual Drude conductivity in the normal state.
The second term gives conventional Josephson current in equilibrium,
while the third term, the anomalous current,\cite{VolkovI} appears in
non-equilibrium superconducting junctions. Another diagonal component
of $\hat{J}^K$,
\begin{equation} \label{currents+}
I_+ \equiv (1/ 4)\mathop{\textrm {Tr}}\hat{J}^K   = D_+\partial f_+ +
I_s f_- + I_{\mathit{an}} \partial f_-,
\end{equation}
has the meaning of the net quasiparticle current (the sum of the
electron and hole probability currents). Explicit equations for the
spectral characteristics of the junction, $D_\pm$, $I_s$, and
$I_{an}$, are conveniently written in terms of the following
parametrization of the matrix $\hat{g}$,
\begin{equation} \label{Green}
\hat{g} = \hat{u} + \hat{v} = \sigma_z u + \exp(i\sigma_z
\psi)i\sigma_y v, \quad u^2-v^2=1,
\end{equation}
The function $u$ is related to the quasiparticle density of states
(DOS), normalized by its value in the normal state, $N(E,x)
=\mathop{\textrm {Re}}u^R(E,x)$, while the function $v$ is related to
the spectral density of the condensate. The complex phase $\psi$
appears in the presence of a supercurrent. In these notations, the
diffusion coefficients read
\begin{eqnarray} \label{D}
D_\pm &=& (1 /4)\mathop{\textrm {Tr}} \left(1-\hat{u}^R \hat{u}^A \mp
\hat{v}^R \hat{v}^A\right)
\\ \nonumber
&=&(1 / 2)\left[1+\left|u^R\right|^2 \mp \left|v^R\right|^2
\cosh\left(2\mathop{\textrm {Im}}\psi^R \right)\right].
\end{eqnarray}
In the normal state, $D_\pm$ turn to unity.

The spectral densities of the supercurrent and ano\-malous current
are given by equations,
\begin{subequations}\label{Spectral}
\begin{eqnarray}
I_s = (1/4)\mathop{\textrm {Tr}}\sigma_z\left(\hat{v}^R
\partial\hat{v}^R\! - \hat{v}^A\partial\hat{v}^A\right) =
\!-\mathop{\textrm {Im}}\left(v^2\partial\psi\right)^R\!\!\!,
\\
I_{\mathit{an}} = (1/4)\mathop{\textrm {Tr}}\sigma_z
{\hat{v}^R}{\hat{v}^A} = -\left|v^R\right|^2 \sinh
\left(2\mathop{\textrm {Im}} \psi^R\right)/2.
\end{eqnarray}
\end{subequations}
In Eqs.~(\ref{D}) and (\ref{Spectral}), the relations $(u,v)^A =
-(u,v)^{R \ast}$ and $\psi^A = \psi^{R\ast}$ are used, which follow
from the general relationship $\hat{g}^A = -\sigma_z \hat{g}^{R\dag}
\sigma_z$.\cite{LO}

Calculation of the electric current in Eq.~(\ref{current}) involves
the two steps: first one has to solve the Usadel equations for the
Green functions $\hat g^{R,A}$, and then to solve the kinetic
equations to find the distribution functions.

\section{spectral functions}

The Green function components of Eq.~(\ref{Keldysh}) represent the
Usadel equations for the spectral functions,\cite{Usadel}
\begin{eqnarray}
2Ev &=& i\hbar {\mathcal D} \left[\partial (u \partial v -v\partial
u)- u v(\partial\psi)^2\right],\label{Usadell}
\\
v^2 \partial\psi &=& I, \label{EqPsi}
\end{eqnarray}
where the spatial constant $I(E)$ is related to the supercurrent
spectral density in Eq.~(\ref{Spectral}), $I_s = -\mathop{\textrm
{Im}} I^R$. In terms of the spectral angle $\theta$ related to the
spectral functions as $u = \cosh\theta$, $v = \sinh\theta$, Eq.~(\ref
{Usadell}) takes the form,
\begin{equation} \label{EqTheta}
\partial^2\theta = (2E/ i\hbar{\mathcal D} )\sinh\theta +
I^2\cosh\theta/\sinh^3\theta.
\end{equation}
The two terms in r.h.s. of Eq.~(\ref{EqTheta}) are related to two
different depairing mechanisms, which provide spatial decrease of
$\theta$ towards the middle of the junction. The first term is
associated with the dephasing between the electron and hole wave
functions at finite energy $E$. The second term describes depairing
caused by the time-reversal symmetry breaking due to supercurrent
flow.

The solution of Eqs.~(\ref{EqPsi}) and (\ref{EqTheta}) reads
\begin{eqnarray}
&\displaystyle x_i = \int_{\theta_0}^{\theta(E,x_i)} {d\theta \over
\sqrt{R(E,\theta)}},\quad x_i = x,y, \label{SolutionTheta}
\\
&\displaystyle \psi(E,x_i) = \psi_0+ I\int_0^{x_i} {dz \over
v^2(E,z)}, \label{SolutionPsi}
\\
&\displaystyle R(E,\theta) \equiv C + (4E/ i\hbar{\mathcal D}
)\cosh\theta - (I/ \sinh\theta)^2, \label{FirstInt}
\end{eqnarray}
where $C(E)$ is the integration constant, and $\theta_0$, $\psi_0$
are the spectral functions at the junction node. The boundary
conditions for Eqs.~(\ref{EqPsi}) and (\ref{EqTheta}) are imposed by
the conservation law for the matrix current in Eq.~(\ref{BoundaryJ}),
\begin{equation} \label{BoundaryTheta}
\left. \partial_x\theta\right|_{x=-0} = \left. \partial_x\theta
\right|_{x=+0} + \left. \partial_y\theta\right|_{y=0},
\end{equation}
and similar equation for the function $\psi$. At the normal
electrode, the current $I$ turns to zero, which means that $I \equiv
0$ along the injection lead, and therefore $\psi(y)\equiv
\text{const}$. Hence, the derivative of $\psi$ is continuous in the
horizontal lead, whereas $\theta(x)$ has a kink at $x=0$. In what
follows, we consider symmetric junction, in which $\theta(x)$ is
even, and $\psi(x)$ is odd function. In this case, the phase $\psi$,
together with the anomalous current $I_{\mathit{an}}$, turns to zero
at the junction node, and the kink in $\theta(x)$ is symmetric,
$\theta_0^\prime \equiv \partial_x\theta|_{x=+0} = -
\partial_x\theta|_{x=-0}$.

The boundary conditions at the NS interfaces depend on the interface
resistance. Below we analyze the two different situations related to
perfect and high-resistive interfaces, respectively.

\subsection{Transparent interfaces}

If the interface electric resistance $R_{\it NS}$ is much smaller
than the normal resistance $R_N$ of the horizontal lead, one can
assume the spectral functions to be continuous at $x = \pm d$,
namely, $\psi(\pm d) = \pm \phi/2$, and $\theta(\pm d) = \theta_S
\equiv {\text{{Arctanh}}}\, (\Delta/E)$; at the normal electrode
$\theta =0$. In this limit, the second term in the r.h.s. of
Eq.~(\ref{FirstInt}) can be neglected due to large gradients of the
spectral functions along the leads, which results in linear change of
$\theta$ along the injection lead, $\theta(E,y) = \theta_0(1-y/L)$,
and the boundary condition in Eq.~(\ref{BoundaryTheta}) takes the
form
\begin{equation} \label{FinalBoundary}
\theta_0 = 2\theta^\prime_0 L.
\end{equation}

The analytical expressions for the spectral functions in the
horizontal lead within this approximation have been found in
Ref.~\onlinecite{Heik}. In the right lead, they are given by
\begin{eqnarray}
u(E,x)& = &\widetilde{u}_0\cosh\left(\alpha + \Lambda x/ d\right),
\label{uRL}
\\
\psi(E,x)& =& \arctan\left[\widetilde{v}_0^{-1} \tanh\left(\alpha +
\Lambda x/ d\right)\right] - p.\label{psiRL}
\end{eqnarray}
The solution in the left lead are obtained by the change of the signs
of $x$, $\phi$, $\alpha$, and $p$. The spatial constants in
Eqs.~(\ref{uRL}) and (\ref{psiRL}) can be parameterized as
\begin{eqnarray}
I \!&=&\! {\widetilde{v}_0\over d} \Lambda, \quad \Lambda =
\mathop{\textrm {Arccosh}} {u_S\over \widetilde{u}_0} -
\mathop{\textrm {Arccosh}} {u_0\over \widetilde{u}_0},\label{Nspsi}
\\
u_0\! &\equiv&\! \cosh\theta_0 = \widetilde{u}_0\cosh\alpha =
u_S(\widetilde{E},\widetilde\Delta),\label{SolutionTilde}
\\
\widetilde{u}_0 \!&=&\! u_S(E,\widetilde\Delta), \; \widetilde{E}\!
=\! E\cos p,\; \widetilde\Delta \!=\! \Delta\cos(\phi/2\!+\!p)\!,
\label{Current}
\end{eqnarray}
where $u_S(E,\Delta) =E/\sqrt{E^2-\Delta^2}$, and expressed via a
single parameter $p$, which is to be evaluated from the equation
following from Eq.~(\ref{FinalBoundary}),
\begin{equation} \label{rt1}
{\widetilde{u}_0} \Lambda \sin p = a\theta_0.
\end{equation}

The magnitude of the parameter $p$ is controlled by the parameter
$a=d/2L$. When $a$ decreases, i.e., the resistance of the injection
lead increases, $p$ turns to zero, according to Eq.~(\ref{rt1}), and
the spectral functions approach their values in a closed short SNS
junctions. In the limit $a=0$, DOS has the proximity gap
$|\Delta_\phi|$, where $\Delta_\phi = \Delta\cos(\phi/2)$, and
reveals a BCS-like singularity at the gap edge,
\begin{equation} \label{DOSZeroa}
N(E,0)  = E \Theta(E-|\Delta_\phi|) / \sqrt{E^2 -\Delta_\phi^2},
\end{equation}
where $\Theta(x)$ is the Heaviside step function. The
super\-cur\-rent spectral density $I_s(E)$ spreads over the region
$|\Delta_\phi| \leq E \leq \Delta$, and has the singularity at the
proximity gap edge as well,\cite{Mats}
\begin{equation} \label{Zeroa}
I_s(E) =  {\pi\over 2d} {\Delta_\phi\Theta(E - |\Delta_\phi|) \over
\sqrt{E^2 - \Delta_\phi^2}}\Theta(\Delta - E).
\end{equation}
\begin{figure}[tb]
\epsfxsize=8.5cm\epsffile{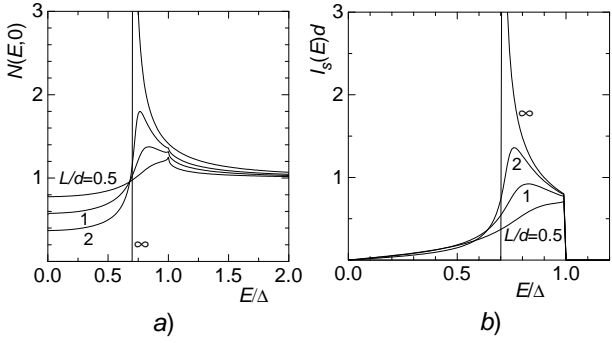} 
\caption{The density of states $N(E,0)$ at the device node (a), and
the supercurrent spectral density $I_s(E)$ at $\phi = \pi/2$ (b), for
several values of the ratio $L/d$, calculated by using numerical
solution of Eq.~(\ref{rt1}).}
\label{NandI}\vspace{-4mm}
\end{figure}

In a general case, $a \neq 0$, the proximity of the normal reservoir
leads to finite DOS at all energies, as shown in Fig.~\ref{NandI}(a),
though it is noticeably suppressed at $E < |\Delta_\phi|$ for $L
\gtrsim d$. The supercurrent spectral density at finite $d/L$ extends
over the whole subgap region [see Fig.~\ref{NandI}(b)], while at
$E>\Delta$, both $I_s$ and $I_{\mathit{an}}$ turn to zero. Thus, in
short diffusive junctions with trans\-pa\-rent interfaces, the
supercurrent is carried exclusively by the bound Andreev states
confined to the potential well formed by the junction. However, this
result is only correct to zero approximation with respect to the
small parameter $d/\xi_0$.\cite{note1}

\subsection{Opaque interfaces}

The effect of the interface becomes important when the interface
resistance $R_{\it NS}$ exceeds the resistance of the normal
conductor $R_N = 2d/\sigma $, $r =R_{\it NS}/ R_N \gg 1$. In
particular, the magnitude of the Josephson current is determined by
the $R_{\it NS}$ rather than by $R_N$ in the limit $r\gg 1$. At the
same time, as we will see below, the suppression of the proximity
effect is governed by much smaller parameter $r(d/\xi_0)^2 \ll r$,
and the proximity effect can be strong even when $r\gg 1$.

A high-resistive interface can be modelled by an effective tunnel
barrier characterized by its resistance $R_{\it NS}$ in the normal
state, which results in the following boundary conditions for the
Green functions\cite{KL} at $x=d$,
\begin{subequations} \label{Boundary1}
\begin{eqnarray}
\sigma  R_{\it NS} \partial \theta_N = u_N v_S \cos(\phi/2-\psi_N) -
u_S v_N, \label{BoundaryTheta1}
\\
\sigma  R_{\it NS} I = v_N v_S \sin(\phi/2 -\psi_N), \quad I = v^2_N
\partial \psi_N, \label{BoundaryPsi1}
\end{eqnarray}
\end{subequations}
and similar for $x=-d$ (the indices $N$ and $S$ refer to the normal
and superconducting sides of the interface).

In the limit $r \gg 1$, the spatial variation of the spectral phase
is strongly non-homogeneous: the phase drops at the barriers and is
small in the normal region, $\psi_N \ll 1$, along with the spectral
current $I$. The spatial variation of the spectral function $u$ is
small and can be approximated by weakly varying parabolic function,
\begin{equation} \label{Parabolic}
u(E,x) \approx u_0 \left[ 1 + (\beta/ 2) \left(x/ d\right)^2\right].
\end{equation}
In Eq.~(\ref{Parabolic}), we neglected the effect of the injection
lead, assuming its resistance to be larger than $R_{\it NS}$, $1/a
\gg r$. The coefficient $\beta \ll 1$ is to be found from
Eq.~(\ref{FirstInt}), in which the electron-hole dephasing effect has
to be taken into account because it now becomes comparable with the
small current-induced depairing,
\begin{equation} \label{beta}
\beta = -2i{E\over \Delta} \left({d\over \xi_0}\right)^2 {v^2_0 \over
u_0} + \left({Id \over v_0}\right)^2.
\end{equation}
In Eq.~(\ref{SolutionPsi}), we may neglect spatial variations of the
integrand which results in a linear spatial dependence of the phase,
$\psi(E,x) \approx I x/v_0^2 \sim r^{-1}$.

By making use of Eqs.~(\ref{Parabolic}) and (\ref{beta}), the
boundary conditions in Eqs.~(\ref{Boundary1}) give the equation for
the spectral functions $u_N$ and $v_N$,
\begin{equation} \label{BoundaryTheta2}
{u_S \over u_N} = {i\gamma E\over u_N \Delta} + {v_S\over
v_N}\cos{\phi\over 2}-r\psi^2_N, \;\, \gamma =  2r \left({d/
\xi_0}\right)^2,
\end{equation}
and the expression for the spectral current $I$,
\begin{equation}
\label{BoundaryPsi2} I=  {v_N v_S \over \sigma  R_{\it NS}} \sin
{\phi \over 2},
\end{equation}
in which we omitted the small phase $\psi_N$ from the trigonometric
functions. Equation (\ref{BoundaryTheta2}) describes three mechanisms
of depairing. The first term in the r.h.s.\ represents the
electron-hole dephasing within the normal metal. The parameter
$\gamma$ determines the magnitude of the energy gap $E_g \sim
{\Delta}/(1+\gamma)$ in the spectrum of the horizontal lead (see
below). The second term describes suppression of the condensate
function $v_N$ due to rapid change of the spectral phase across the
tunnel barrier. This effect is similar to the mechanism which
produces the Andreev bound states in the vicinity of the tunnel
junction in the ballistic \cite{Fur,Wendin} as well as diffusive
\cite{BBG} Josephson structures. The third term is caused by the
supercurrent flow through the normal lead. Neglecting this small
($\sim r^{-1}$) term, we obtain the solution of
Eq.~(\ref{BoundaryTheta2}),
\begin{equation} \label{vN}
v_N = {\widetilde\Delta \over \sqrt{E^2\! -\!
\widetilde\Delta^2}},\;\; \widetilde\Delta(E,\phi) =
{\Delta_\phi\over 1+\gamma \sqrt{\Delta^2\! -\! E^2}/\Delta}.
\end{equation}
According to Eq.~(\ref{vN}), the energy gap $E_g(\phi)$ in the
spectrum of the junction is to be determined by the equation $E_g =
|\widetilde\Delta(E_g,\phi)|$, whose solution can be well
approximated by a simple relation $E_g = |\Delta_\phi| / (1+\gamma)$.

\begin{figure}[tb]
\epsfxsize=8.5cm\epsffile{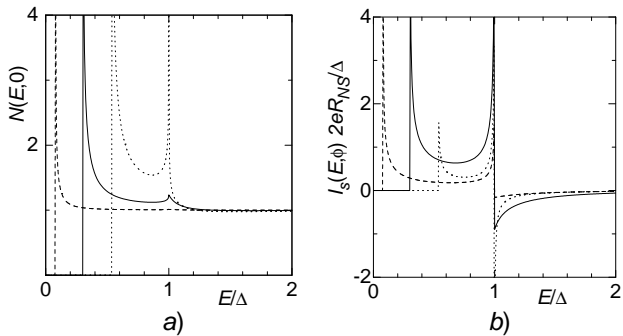} 
\caption{Variations in the DOS (a), and in the supercurrent spectral
density (b) with the phase difference in SINIS junction, calculated
from Eqs.~(\ref{vN}) and (\ref{IS}) at $\gamma = 1$: $\phi=0.1\pi$
(dotted lines), $0.6\pi$ (solid lines), and $0.9\pi$ (dashed lines).}
\label{jsg} \vspace{-4mm}
\end{figure}

Thus, the regime of strong proximity effect with the proximity gap
being of the order of $\Delta$ [Fig.~3(a)], persists in short
junction despite of high-resistive interfaces, $R_{\it NS} \gg R_N$,
as long as the condition $\gamma \lesssim 1$ holds. In this case, the
supercurrent spectral density,
\begin{equation} \label{IS}
I_s(E,\phi) = -\mathop{\textrm {Im}}I=-{\sin(\phi / 2)\over \sigma
R_{\it NS}} \mathop{\textrm {Im}} \left[ {\Delta
v_N(E,\phi)\over\sqrt{E^2 - \Delta^2}} \right],
\end{equation}
extends over all quasiparticle states above the proximity gap,
including the continuum states above the bulk energy gap $\Delta$,
where $I_s(E)$ is negative [Fig.~3(b)]. In the limit $\gamma \ll 1$
(recall that we nevertheless assume here the interface resistance to
be large, $r \gg 1$), the dephasing effect becomes negligibly small
and the energy gap approaches $|\Delta_\phi|$, similar to the perfect
SNS junction discussed above. However, the shape of the supercurrent
spectral density essentially differs from Eq.~(\ref{Zeroa}),
\begin{equation} \label{gammasmallI}
I_s(E) =  {\sin\phi\over 2\sigma R_{\it NS}} {\Delta^2\Theta(E -
|\Delta_\phi|)\Theta(\Delta - E) \over \sqrt{E^2 -
\Delta_\phi^2}\sqrt{\Delta^2 - E^2}}.
\end{equation}
It is interesting to note that the expression for the equilibrium
Josephson current obtained from Eq.~(\ref{gammasmallI}),
\begin{equation} \label{gammasmall}
j_s = {\Delta \sin\phi \over 2eR_{\it NS}} K\big(\left|\sin (\phi /
2) \right|\big),\quad \gamma \ll 1 \quad (T=0)
\end{equation}
($K$ is the elliptic integral), can be reproduced with the arguments
of the scattering theory, similar to the case of perfect diffusive
SNS junction,\cite{Been} by applying the transmissivity distribution
for a normal symmetric double-barrier structure.\cite{deJong}  Such a
possibility is explained by the absence of electron-hole dephasing in
this limit.

The proximity gap is strongly suppressed, $E_g \ll \Delta$, only in
the limit of very large barrier strength, $\gamma \gg
1$.\cite{VolkovSINIS} This is the effect of enhanced electron-hole
dephasing, similar to the case of a long diffusive SNS junction,
where the proximity gap is also reduced due to the dephasing effect
and closes at $\phi = \pi$. This situation is qualitatively different
from the case of the tunnel junction with a single barrier, giving
the Josephson current in junctions with two strong barriers,
\begin{equation} \label{gammalarge}
j_s = {\Delta \sin\phi \over 2eR_{\it NS}\gamma} \ln{4\gamma \over
|\cos(\phi/2)|},\quad \gamma \gg 1 \quad (T=0),
\end{equation}
to be much smaller than the result of the tunnel model.\cite{AB}


\section{Kinetic equations}

In the absence of inelastic collisions, the kinetic equations in each
lead have the form of conservation laws for the spectral currents
$I_\pm(E)$,
\begin{equation} \label{Kinetic}
D_\pm\partial f_\pm + I_s f_\mp \pm I_{\mathit{an}} \partial f_\mp
\equiv I_\pm(E) = \mbox{const}.
\end{equation}

At the junction node, the conservation law for the matrix currents in
Eq.~(\ref{BoundaryJ}) imposes the boundary condition
\begin{equation} \label{BoundaryI}
I_\pm^L = I_\pm^R + I_\pm^V,
\end{equation}
where the indices $L$, $R$, and $V$ refer to the left, right, and
injection leads, respectively. At the transparent interfaces, the
distribution functions are determined by the local-equilibrium
population in the reservoirs,
\begin{subequations} \label{BoundaryF}
\begin{eqnarray} \label{BoundaryFNS}
\displaystyle f_+(\pm d) = \tanh {E\over 2T} \quad (E>\Delta),\quad
f_-(\pm d) = 0,
\\  \label{n} \displaystyle f_\pm(L)  =  n_\pm \equiv{1\over
2}\left[ \tanh{E+eV \over 2T} \pm \tanh{E-eV \over 2T} \right]\!\!.
\end{eqnarray}
\end{subequations}
At $E<\Delta$, the quasiparticle population in the leads is
disconnected from the superconducting reservoirs due to complete
Andreev reflection, and the quasiparticle density function $f_+$ is
determined by the condition of the absence of the net probability
current, $I_+=0$. Due to the conservation law in
Eq.~(\ref{BoundaryI}), the subgap probability current $I_+$ turns to
zero within the entire device.

In the energy region $E > {\Delta}$, where the currents $I_s$ and
$I_{\mathit{an}}$ turn to zero, and the diffusion coefficient $D_+$
turns to unity, the kinetic equations have a simple solution,
\begin{equation} \label{IE>D}
I_+^R = {n_0 - n_+ \over R_+ + 2R_+^V}, \quad I_-^R = -{n_- \over R_-
+ 2R_-^V}.
\end{equation}
Here the quantities $R_+ = d$, $R_+^V = L$, $R_-=d\langle 1/D_-^{R,L}
\rangle$, $R_-^V=L\langle 1/D_-^V \rangle$ play the role of effective
resistances of the leads for the spectral currents $I_\pm$, and the
angle brackets denote spatial averaging along the leads. The currents
in the left lead are equal by magnitude but flow in opposite
directions, $I_\pm^L = -I_\pm^R$, and therefore the currents in the
injection lead are twice the currents in the left lead. Combining
this result with the relation $f_+(0) -n_0= -R_+ I_+^R$, following
from the kinetic equations, we find that in the limit of long
injection lead the boundary condition at the junction node becomes
independent of applied voltage,
\begin{equation} \label{equilf}
f_+(0)=n_0  , \quad L\gg d,
\end{equation}
which implies that the quasiparticles in horizontal leads are in
equilibrium with the superconducting reservoirs.

Within the subgap energy region, $E < {\Delta}$, the situation is
more complex due to appearance of the currents $I_s$ and
$I_{\mathit{an}}$; the only simplification is due to the zero
quasiparticle current, $I_+ = 0$. By this reason, $f_+^V =
{\text{const}} = n_+$ within the entire injection lead, including the
junction node. Thus, the boundary conditions for the distribution
functions in the horizontal leads read,
\begin{equation} \label{BoundaryE<D}
f_+(0) = n_+, \;\; f_-(0) = n_- - R_-^V I_-^V, \;\; f_-(\pm d) = 0.
\end{equation}
Taking advantage of the symmetry of the quantities $D_\pm(x)=
D_\pm(-x)$, $ I_{\mathit{an}}(x)= -I_{\mathit{an}}(-x)$, we separate
the even and odd parts of the distribution functions, $f^{s,a} (x) =
[f (x) \pm f (-x)]/2$, in Eqs.~(\ref{Kinetic}), which then become
split in the two independent pairs of kinetic equations. One pair
that couples $f_+^s$ and $f_-^a$,
\begin{subequations} \label{Kinetic1}
\begin{eqnarray}
D_+\partial f_+^s + I_s f_-^a - I_{\mathit{an}} \partial f_-^a & = &
0,
\\
D_-\partial f_-^a + I_s f_+^s + I_{\mathit{an}} \partial f_+^s & = &
\left(I_-^R + I_-^L\right)/2,
\end{eqnarray}
\end{subequations}
has a constant solution, $ f_+^s= n_+$, $ f_-^a = 0$, consistent with
the boundary conditions, which yields the relation $I_s n_+ =
\left(I_-^R + I_-^L \right)/2 $. As we will see later,
Eq.~(\ref{currentJdef}), the non-equilibrium Josephson current $j_s$
has the form $j_s = \left(j^R + j^L \right)/2$, and taking into
account Eq.~(\ref{current}), we arrive at the following
result,\cite{note2}
\begin{equation} \label{currentJ}
j_s = {\sigma_N\over e}\int_0^\Delta dE\,I_s(E)n_+(E).
\end{equation}

The second pair of kinetic equations couples the functions $f_+^a$
and $f_-^s$,
\begin{subequations} \label{Kinetic2}
\begin{eqnarray}
D_+\partial f_+^a + I_s f_-^s - I_{\mathit{an}} \partial f_-^s & = &
0,
\\
D_-\partial f_-^s + I_s f_+^a + I_{\mathit{an}} \partial f_+^a & = &
-I_-^V/2.
\end{eqnarray}
\end{subequations}
Since the source term and also the boundary conditions to these
equations, Eq.~(\ref{BoundaryE<D}), depend on $I_-^V$, these
functions determine the dissipative current.  The solution to
Eq.~(\ref{Kinetic2}) in general case must be found numerically.

At zero temperature, it is possible to further extend the analysis.
By making use of a step-wise shape of the distribution functions
$n_\pm= \Theta[\pm(E - eV)]$, we find that a trivial solution, $f_+
^a= 0$, $f_-^s = 0$, $I_-^V = 0$, satisfies Eq.~(\ref{Kinetic2}) and
all the boundary conditions at $E > eV$. Thus the dissipative current
vanishes in this energy interval. On the other hand, at $E < eV$,
where $n_+ = 0$ and $n_-=1$, equations (\ref{Kinetic2}) have a
non-trivial solution, which implies that the dissipative current
exists at these energies, while the Josephson current is zero,
according to Eq.~(\ref{currentJ}). Thus at zero temperature the
dissipative and non-dissipative currents flow within the separate
energy regions, which do not overlap: The injection current spreads
over the energy region $0<E<eV$,
\begin{equation} \label{currents0}
j^V = {\sigma_N\over e}\int_0^{eV} dE\,I^V_-(E),
\end{equation}
while the supercurrent occupies the region $eV<E<\Delta$,
\begin{equation} \label{currents1}
j_s =\Theta(\Delta-eV){\sigma_N\over e}\int_{eV}^\Delta dE\,I_s(E).
\end{equation}

The analysis for the subgap region also applies to the case of
resistive interfaces, $r\gg 1$. However, at the energies $E> \Delta$,
the supercurrent $I_s$ and anomalous current $I_{an}$ are nonzero and
give additional contribution to the Josephson current in
Eq.~(\ref{currents1}).


\section{Nonequilibrium Josephson current}

In equilibrium, the Josephson current is given by the second term in
Eq.~(\ref{currents}), as it was mentioned in Sec.~II. Under
non-equilibrium conditions, this connection becomes ambiguous and
needs reconsideration. The reason is that the appearance of the
dissipative currents and related gradients of the distribution
functions in Eq.~(\ref{currents}) will lead to spatial variation of
the supercurrent term along the horizontal lead. To find an
appropriate equation for the observable non-equilibrium Josephson
current, we refer to a generic definition of the dc Josephson effect,
as a current flow through a junction without any dissipation. In our
case, the rate of the energy transfer from a voltage source to the
junction is given by equation,
\begin{equation}
W = \int_{-d}^d dx\; j(x){dV\over dx} = j^LV^L + j^RV^R,
\end{equation}
where $V^{L,R}$ are the voltage drops at the left/right leads. The
stationary Josephson effect assumes zero voltage drop between the
superconducting electrodes, $V^L +V^R=0$, thus the non-dissipative
current component must satisfy the equation $j^L_s = j^R_s \equiv
j_s$. Combining this equation with the Kirchhoff's rule, we arrive at
the following definition of the Josephson current through the
currents in the left, right, and injection leads,
\begin{equation}\label{currentJdef}
j^{L,R} = j_s \pm j^V/2.
\end{equation}
Thus, in order to evaluate observable Josephson current in a general
case, it is necessary to calculate the injection current and the
current in one of the horizontal leads, and then apply
Eq.~(\ref{currentJdef}). In particular case of symmetric junction,
this procedure leads to Eq.~(\ref{currentJ}).

Persistent current in a SQUID is the most fundamental manifestation
of the Josephson effect. The Josephson current in
Eq.~(\ref{currentJdef}) coincides with the circulating current, and
it can be directly measured by measuring the induced flux with an
external magnetometer.

\begin{figure}[tb]
\epsfxsize=8.5cm\epsffile{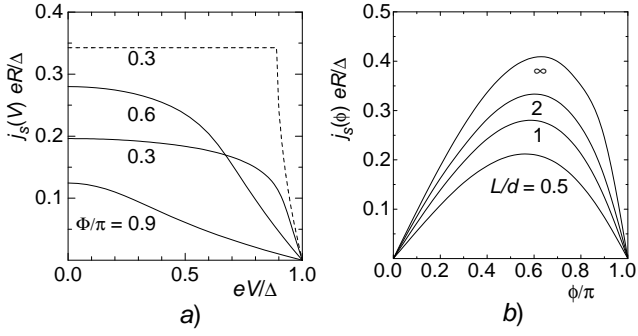} 
\caption{Josephson current vs voltage at different phases: solid
lines - $L=d$, dashed line - $L \gg d$ (a), and Josephson current vs
phase at several values of the ratio $L/d$ (b).}
\label{EquilCurrent} \vspace{-4mm}
\end{figure}

Let us first consider junctions with transparent interfaces, where
the non-equilibrium Josephson current is given by
Eq.~(\ref{currents1}). Since the spectral density $I_s$ is positive
in this case, as it is found from numerical solution of
Eq.(\ref{rt1}), and the population of the subgap states is depleted
with increasing voltage, the injection will suppress the Josephson
current and block it completely at $eV > \Delta$ [see
Fig.~\ref{EquilCurrent}(a)]; however, the current direction cannot be
reversed. The Josephson current weakly depends on the applied voltage
and is close to the equilibrium value as long as the voltage is
smaller than the proximity gap value, $eV<|\Delta_\phi|$. We note
that this equilibrium value differs from that in closed SNS
junctions,\cite{KO} it is reduced due to the proximity of the normal
reservoir and therefore depends on the length of the injection lead,
as shown in Fig.~\ref{EquilCurrent}(b). At larger voltage, the
Josephson current-voltage dependence $j_s(V)$ becomes more steep,
especially for the small phase differences.

For a long injection lead, Eq.~(\ref{currents1}) takes the form,
\begin{equation} \label{js_zeroa}
j_s(\phi,V) = {\pi\Delta_\phi \over 2eR_N}
\int_{|\Delta_\phi|}^\Delta dE\; {n_+(E) \over \sqrt{E^2 -
\Delta_\phi^2}}.
\end{equation}
At zero temperature, the integration in Eq.~(\ref{js_zeroa}) can be
explicitly performed,\cite{Heik}
\begin{eqnarray} \label{kink}
j_s(\phi,V) &\!=\!& {\pi \Delta_\phi \over 2eR_N} \ln{1+\sin(\phi/2)
\over f(V)\! + \!\sqrt{f^2(V)\!-\!\cos^2(\phi/2)}},
\\
f(V) &\!=\!& {\text{max}}[eV/\Delta, \cos(\phi/2)], \;\; eV
> |\Delta_\phi|.
\end{eqnarray}

This current-voltage dependence is shown in
Fig.~\ref{EquilCurrent}(a) by a dashed line. In this case, the
Josephson current at $eV\!<\!|\Delta_\phi|$ is constant and equal to
the equilibrium value.

To estimate the efficiency of the Josephson transistor let us
consider the most steep part of the current-voltage characteristic,
$j_s(\phi,V)$, at small phase and at large voltage. For example, at
$\phi = 0.3\pi$, when the equilibrium Josephson current approaches
about of $0.7$ of its critical value $j_c$, the switching effect
occurs within a small voltage interval $\delta V \sim 0.1\Delta/e$.
The current gain in this case, $\delta j_s/\delta j^V \sim
0.7j_c/G\delta V \sim 7(L/d)$, exceeds unity even for comparable
lengths of the leads, and it can be further enhanced by making the
injection lead longer. The upper bound for the gain is imposed by the
condition of small quasiparticle dwelling time $L^2/{\mathcal D}$,
compared to the quasiparticle relaxation time $\tau$, $L^2/{\mathcal
D}\ll\tau$.

\begin{figure}[tb]
\epsfxsize=8.5cm\epsffile{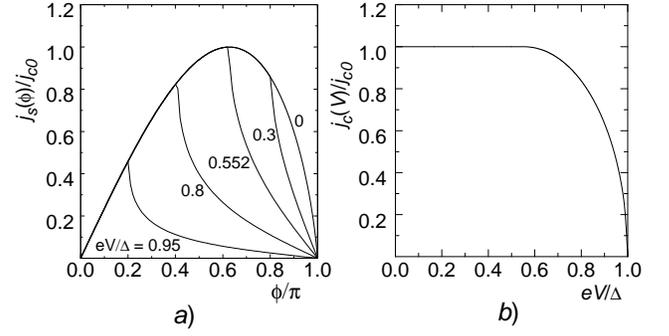} 
\caption{Current-phase relations from Eq.~(\ref{kink}) at different
voltage (a), and critical current vs voltage (b) at $L \gg d$. For a
given voltage, the current-phase relation follows the equilibrium law
as soon as the proximity gap edge is above the energy $eV$, but it is
significantly suppressed when the gap edge is below $eV$.}
\label{j(phi)}\vspace{-4mm}
\end{figure}

The non-equilibrium Josephson current-phase dependence for the
junctions with transparent interfaces and high-resistive injection
lead is shown in Fig.~\ref{j(phi)}(a) for different applied voltages.
The kinks on the graphs correspond to the phase values, at which the
applied voltage equals the proximity gap, $\phi_0(V) =
2\arccos(eV/{\Delta})$. At smaller phases, $0\! < \!\phi\! <\!
\phi_0(V)$, the current-phase dependence has an equilibrium form,
while at larger phases it is considerably distorted. Correspondingly,
the critical current $j_c(V)$ remains independent of applied voltage
until $\phi_0(V)$ exceeds the value $\phi_m = 1.97$, at which the
equilibrium supercurrent approaches its maximum value $j_{c0} =
j_c(0) = 0.66\pi\Delta/2eR$. At larger voltage, $eV\! > \!\Delta
\cos(\phi_m/2) = 0.55\Delta$, the critical current decreases and
turns to zero at $eV \geq \Delta$, as shown in Fig.~\ref{j(phi)}(b),
\begin{equation} \label{jc}
{j_c(V) \over j_c(0)} = {j_s[\phi_0(V),0] \over j_c(0)} = 1.51
{eV\over\Delta} \mathop{\textrm {Arccosh}} {\Delta\over eV}.
\end{equation}

\begin{figure}[tb]
\epsfxsize=8.5cm\epsffile{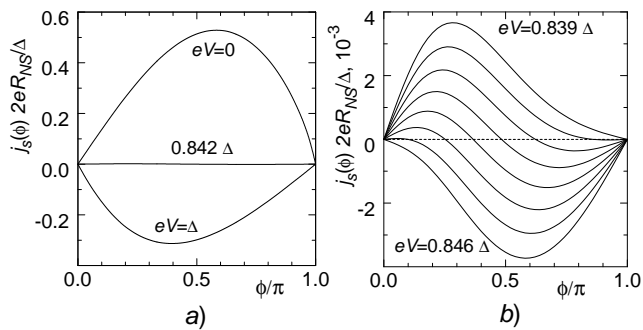} 
\caption{Inversion of the current-phase relation $j_s(\phi)$ under
applied voltage in a SINIS junction at $\gamma = 1$. The voltage step
between the curves in right panel is $10^{-3}\Delta/e$.}
\label{inverse}\vspace{-4mm}
\end{figure}

In junctions with resistive interfaces, $r\gg 1$, the cur\-rent-phase
dependence is more interesting, because of the possibility of the
Josephson current inversion and the crossover to the $\pi$-junction
regime. This results from the negative contribution of the energies
$E > \Delta$ to the Josephson current which turns to zero before the
voltage achieves the gap value, $eV \lesssim \Delta$, when the
positive and negative parts of $I_s(E)$ compensate each other. At
larger voltage the current becomes negative, as shown in
Fig.~\ref{inverse}(a). Detailed analysis of the crossover region can
be made for the junctions with high-resistive injection lead, $L/d\gg
r\gg 1$, and at zero temperature. In this case, in the horizontal
leads, the distribution function $f_-$ is small, and the function
$f_+$ is approximately constant and approaches the equilibrium value
$n_0(E)=1$ in the superconducting reservoir [see Eq.~(\ref{equilf})].
By these reasons, the small dissipative and anomalous components can
be omitted from the current spectral density, $I_-\approx I_s$, which
then becomes independent of applied voltage at $eV > \Delta$. This
results in the following modification of Eq.~(\ref{currents1}),
\begin{equation} \label{jSINIS}
j_s = {\sigma \over e}\int_{{\mathit {min}}(eV,\Delta)}^\infty
dE\,I_s(E),
\end{equation}
where $I_s(E)$ is to be found from Eqs.~(\ref{IS}) and (\ref{vN}).

As follows from Eq.~(\ref{jSINIS}), the critical voltage, at which
the current turns to zero, depends on the phase, and therefore the
crossover extends over a certain, in fact rather small, voltage
interval, as shown in Fig.~\ref{inverse}(b). When the voltage
approaches the critical region, a new current node in the
current-phase dependence splits from the node at $\phi = \pi$, then
it moves towards smaller $\phi$; the process ends when the extra node
approaches $\phi = 0$. Such a fine structure of the Josephson current
inversion has been observed experimentally in long SNS
junctions.\cite{Baselmans02} At very large interface resistance,
$\gamma \gg 1$, this fine structure becomes irresolvable because of
in this limit the phase dependence in $I_s(E,\phi)$ for relevant
energies is given by a prefactor $\sin\phi$, and therefore the
compensation effect appears simultaneously at all
phases.\cite{VolkovSINIS}

\section{Critical current in current biased junction}

In experiment, the current bias setup is often employed for
investigation of the dc Josephson current, Fig.~\ref{device}(c). In
equilibrium, the maximum value of the current flowing through the
junction without creating a voltage drop coincides with the maximum
current in the current-phase dependence. This is not the case for a
non-equilibrium junction with current injection: the ``critical''
current is contributed by both the non-equilibrium Josephson current
and the injection current.

Suppose the voltage is applied between the injection elec\-trode and
left superconducting electrode, Fig.~\ref{device}(c), then the
external transport current $j_T$ is equal to the cur\-rent $j^R$ in
the right lead. In this case, the problem of the critical current
evaluation is reduced to the analysis of the phase dependence of $j^R
= j_s - j^V/2$ at given voltage. The requirement of zero potential
difference between the superconducting electrodes is automatically
fulfilled in our calculation (time-independent phase difference). For
simplicity, we consider the junction with perfect interfaces, where
the currents are given by Eqs.~(\ref{currents0}), (\ref{currents1}).

\begin{figure}[tb]
\epsfxsize=8.5cm\epsffile{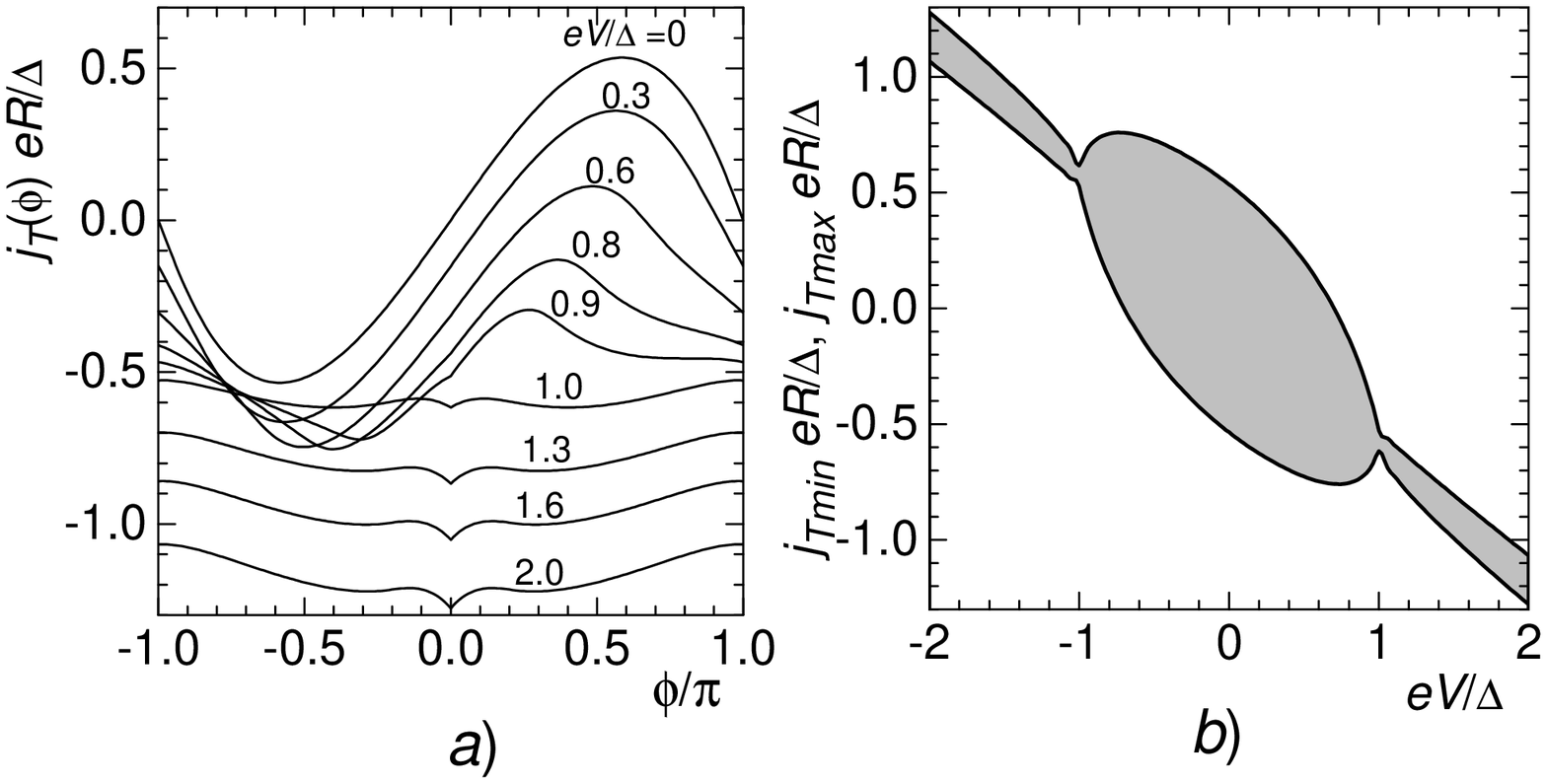} 
\epsfxsize=8.5cm\epsffile{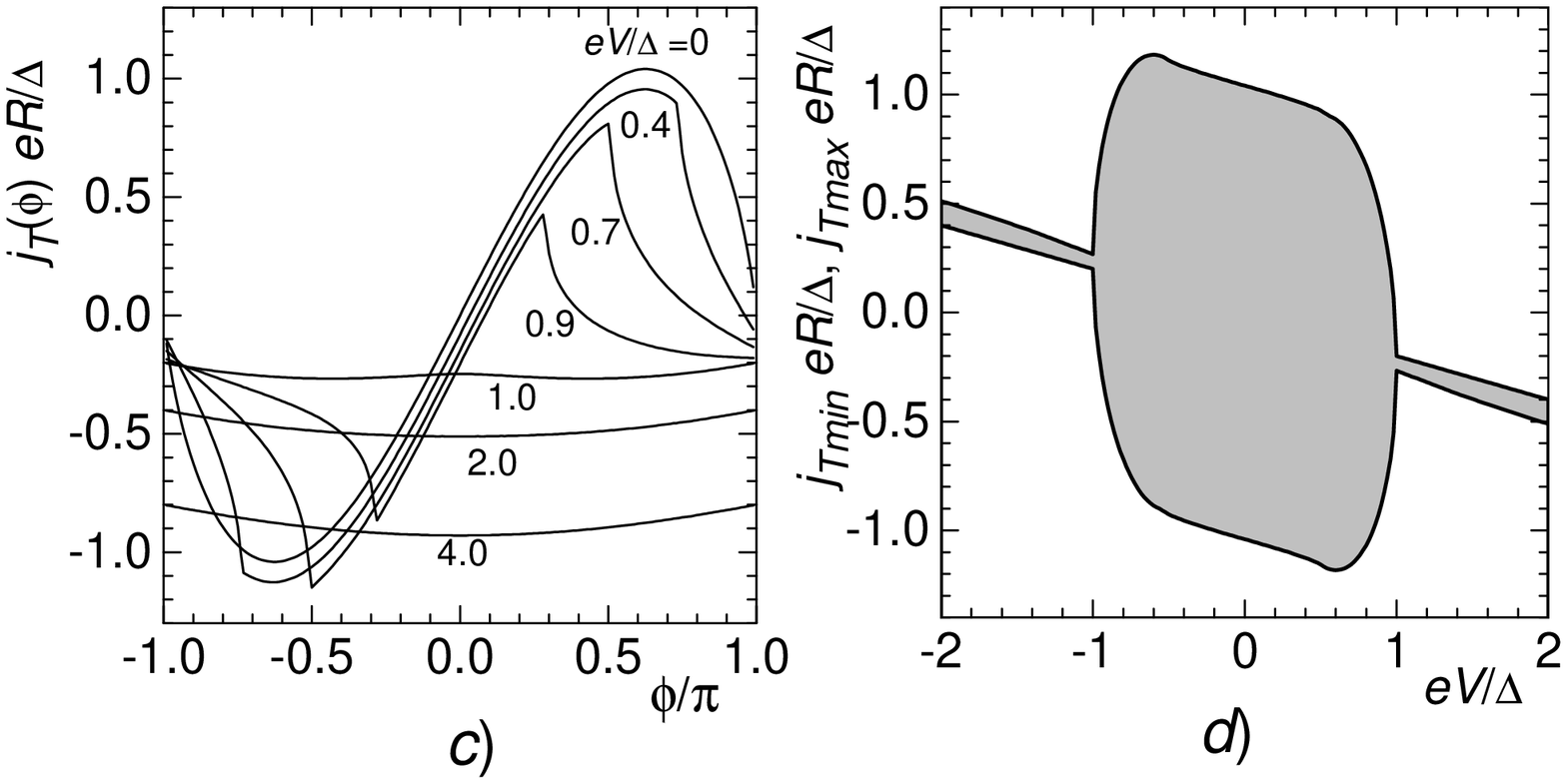} 
\caption{Current $j_T$ vs phase $\phi$ at different injection voltage
(a,c), and critical transport current $j_{\it Tmin}$, $j_{\it Tmax}$
vs voltage (b,d), at $L=d$ (a,b), and $L=5d$ (c,d). Shaded regions
correspond to zero potential difference between the superconducting
electrodes. In oval-like regions, the Josephson current coexists with
the normal current; in shaded stripes at $eV > \Delta$, the Josephson
current is absent (Josephson-like regime). }
\label{Imm}\vspace{-4mm}
\end{figure}

The numerical results of such analysis are shown in
Fig.~\ref{Imm}(a,c). They are obtained by solving numerically
Eq.~(\ref{rt1}) for the spectral functions and Eq.~(\ref{Kinetic})
for the distribution functions, which determine the magnitude of the
injection current. At $eV < \Delta$, when the supercurrent is allowed
to flow through the junction, the current-phase relations are similar
to that depicted in Fig.~\ref{j(phi)}(a). At these voltages, the
Josephson current coexists with the normal current flowing out of the
injection lead. At larger voltages, $eV>\Delta$, the supercurrent is
blocked, however, the transport current still flows through the
junction without voltage drop across it, within a certain range of
the current magnitudes determined by the amplitude of the dependence
$j^R(\phi)$. The existence of such Josephson-like regime without real
Josephson current has been first pointed out for a 4-terminal SNS
junction with opaque interfaces.\cite{VP}

To understand this phenomenon, it is important to remember that the
injection current in NS interferometers is not uniquely determined by
the bias voltage, but also depends on the superconducting phase. In
principle, a similar regime with zero voltage drop across the
junction may appear even for normal reservoirs, at the transport
current $j_T = j^V(V)/2$. This value is unique for the given
injection voltage, and therefore the corresponding dependence
$j_T(V)$ is represented by a straight line. In the superconducting
junctions, such line broadens to a stripe, $j_T = j^V(V,\phi)/2$, due
to the presence of the free parameter $\phi$: The phase adjusts the
injection current for given injection voltage and transport current
to provide zero voltage drop across the junction. The width of the
stripe is determined by the amplitude of the injection current
oscillation with the phase. At $eV < \Delta$, this effect is hidden
by the presence of the true supercurrent [large shaded regions in
Fig.~\ref{Imm}(b,d)], however, it is fully revealed at $eV > \Delta$,
where the supercurrent is suppressed [shaded stripes in
Fig.~\ref{Imm}(b,d)]. In fact, at large voltage, the width of the
shaded stripes is determined by the amplitude of phase oscillations
of the excess injection current. The qualitative difference between
the phase dependence of the excess current ($eV > \Delta$) and the
Josephson current ($eV < \Delta$), is clearly seen in
Fig.~\ref{Imm}(a,c). It is interesting that the ``critical current''
has different sign for positive and negative voltages [the shaded
stripes in Fig.~\ref{Imm}(b,d) are differently oriented with respect
to the straight line]. This is consistent with the fact that the
excess current changes sign along with the applied voltage.

%

\section{Interferometer effect}

\begin{figure}[tb]
\epsfxsize=8.5cm\epsffile{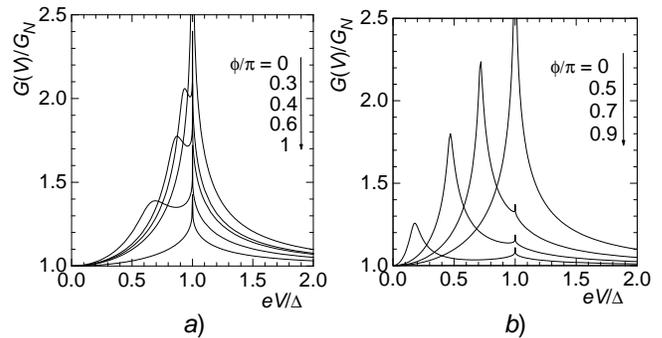} 
\caption{Differential conductance $G$ vs voltage at $L=d$ (a), and $L
= 5d$ (b) for different phases.}
\label{FigeV}\vspace{-4mm}
\end{figure}

In this Section, we investigate the conductance of the injection lead
as a function of the bias voltage and superconducting phase focusing
on its properties due to the strong proximity effect. At small
temperatures, $T \ll eV$, the overall voltage dependence of the
differential conductance is given by Eq.~(\ref{currents0}),
$G(V,\phi) = dj^V/dV = \sigma I_-^V$, where the injection current
$I_-^V$ is to be calculated by numerical solution of Eq.~(\ref{rt1})
for the spectral functions and the kinetic equations (\ref{Kinetic}).
As shown in Fig.~\ref{FigeV}, the conductance has two peaks, at
voltages $eV = {\Delta}$ and $eV \approx {\Delta}_{\phi}$. The first
peak is associated with enhanced transmissivity of the junction due
to the DOS peak at the bulk gap edge. This peak has a logarithmic
singularity at $\phi=0$, and it becomes smeared and decreases while
$\phi$ departs from zero. The second peak manifests a rapid change in
the spectral functions in the vicinity of the spectrum edge in a
short SNS junction (see Fig.~\ref{NandI}), and it can be interpreted
as a resonance transmission due to enhanced DOS at the proximity gap
edge $|\Delta_\phi|$. As soon as $L$ increases, this resonance
becomes more sharp because of the singularities in the spectral
functions at $E = |\Delta_\phi|$ become more pronounced, whereas the
peak at the bulk gap edge, $eV = \Delta$, decreases and vanishes at
$d/L \rightarrow 0$, as shown in Fig.~\ref{FigeV}(b). Furthermore,
the conductance exhibits the reentrance effect: $G(V)$ approaches the
value $G_N =\sigma /(d/2+L)$ in the normal state both at small and
large voltages, as it was predicted for NS point contacts.\cite{AVZ}
We notice that the differential conductance $G(V)$ deviates from
$G_N$ in the short-arm interferometers at the characteristic energy
of the order of ${\Delta}$, in contrast to the long-arm SNS junctions
($d\gg \xi_0$), where the conductance peak appears at the Thouless
energy $E_{\text{Th}}= \hbar {\mathcal{D}}/(2d)^2 \ll
\Delta$.\cite{Reentrance}

At zero phase difference, $\phi=0$, the function $G(V)$ can be found
analytically. In this case, $I_s = I_{\mathit{an}} = 0$, and the
function $f_-$ obeys a simple equation $D_-\partial f_- = I_-$ in
each lead, with the diffusion coefficients
\begin{subequations} \label{DatPhi=0}
\begin{eqnarray}
D_-^{R,L} &=& \cosh^2\left[\mathop{\textrm {Re}}\theta_S(E){1+a|x|/ d
\over 1+a} \right],
\\
D_-^V &=& \cosh^2\left[\mathop{\textrm {Re}}\theta_S(E) {1-y/L \over
1+a} \right]. \label{D1}
\end{eqnarray}
\end{subequations}

From Eqs.~(\ref{currents0}) and (\ref{DatPhi=0}), we obtain
\begin{eqnarray}
&\displaystyle G(V) = G_N \eta(eV) \quad (\phi=0), \label{GatPhi=0}
\\
&\displaystyle \eta(E) = z\mathop{\textrm {Arctanh}}z^{-1}, \quad z=
(E/\Delta)^{{\text{sign}}(E-\Delta)}. \label{Defeta}
\end{eqnarray}

The oscillations of the conductance peak at $eV = \Delta$ with the
phase can be found from the following arguments. At this energy, the
diffusion coefficient $D_-$ turns to infinity in the horizontal lead
which therefore becomes non-resistive with respect to the normal
current, $R_-(\Delta) = 0$. Thus, the differential conductance at
$eV=\Delta$ is completely determined by the resistance of the
injection lead $R_-^V(\Delta) = L\tanh\theta_0/ \theta_0$, where
$\theta_0 = \ln \cot (\phi/4)$,
\begin{eqnarray} \label{peak}
&\displaystyle G_{\text{max}}(\phi) \equiv G(\Delta/e,\phi)= G_N {1+a
\over \cos (\phi/2)}\ln\cot{\phi\over4}.
\end{eqnarray}

According to Eq.~(\ref{peak}), the peak height approaches $G_N(1+a)$
at $\phi=\pi$. At this point, like at $\phi = 0$, the spectral
densities of the superconducting and anomalous current turn to zero,
and only the resistances $R_-$ and $R_-^V$ are involved in the
calculation. Since the condensate function becomes completely
suppressed in the middle of the junction, $\theta_0 = 0$, the
injection lead behaves as a normal wire, and therefore the resistance
$R^V_-$ approaches its normal value $L$. Correspondingly, the
resistance of each horizontal lead coincides with the resistance of a
short NS junction with the length $d$, $R_- = d\tanh(\mathop{\textrm
{Re}} \theta_S)/\mathop{\textrm {Re}}\theta_S$, and therefore $G(V)$
at $\phi = \pi$ is given by
\begin{equation} \label{Gpi}
G(V) = G_N {1+a \over 1+a/\eta(eV)}      \qquad (\phi = \pi) .
\end{equation}

In the limit of long injection lead, $d/L\rightarrow 0$, its
resistance $R_- ^V = L\, {\tanh\mathop{\textrm {Re}}\theta_0
/\mathop{\textrm {Re}}\theta_0}$, $\theta_0 = \mathop{\textrm
{Arctanh}}(\Delta_\phi/ E)$, completely determines the injection
current, and the voltage dependence of the differential conductance
shown in Fig.~\ref{FigeV}(b) is approximately described by
Eq.~(\ref{GatPhi=0}), with $|\Delta_\phi|$ substituted for $\Delta$
in the function $\eta(eV)$.

\begin{figure}[tb]
\epsfxsize=8.5cm\epsffile{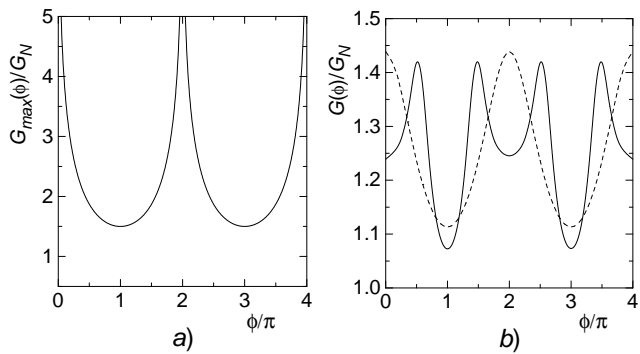} 
\caption{Differential conductance vs phase at $L=d$ and different
voltage: $V=\Delta/e$ (a), $V=0.7\Delta/e$ (b, solid line), and
$V=1.2\Delta/e$ (b, dashed line). }
\label{Gmax}\vspace{-4mm}
\end{figure}

The differential conductance exhibits full-scale $2\pi$-periodic
oscillations with the phase difference $\phi$ (Andreev interferometer
effect). The form of the oscillations is qualitatively different for
the subgap bias region, $eV < \Delta$, and for $eV > \Delta$, as
shown in Fig.~\ref{Gmax}(b). In the latter case, the phase dependence
of $G$ has a comparatively simple form, with maxima at $\phi = 2\pi
n$ and minima at $\phi = (2n+1)\pi$. At $eV < \Delta$, the
differential conductance approaches minima both at even and odd
multiples of $\pi$, which reflects the interplay between the position
and amplitude of the resonance at $eV = |\Delta_\phi|$.

\section{Summary}

We have developed a theory of the non-equilibrium Josephson effect in
a three-terminal diffusive interferometer with short SNS junction
having the length $2d$ much smaller than the superconducting
coherence length $\xi_0$. We focused on the case of strong proximity
effect, when the proximity energy gap in the normal region is of the
order of $\Delta$. For the junction with transmissive NS interfaces,
the density of states $N(E)$ and the supercurrent spectral density
$I_s(E)$ extend over the whole subgap region, $0< E < \Delta$, due to
the proximity to a normal reservoir, and exhibit a considerable
enhancement at the energy equal to the proximity gap $|\Delta_\phi| =
\Delta |\cos (\phi/2)|$ in the spectrum of a closed SNS junction. The
supercurrent spectral density is positive at all relevant energies.
We demonstrated a possibility of the strong proximity effect in a
junction with opaque interfaces whose resistance $R_{\it NS}$ is much
larger than the normal resistance $R_N$ of the junction arms. In such
case, the suppression of the proximity gap, $E_g(\phi) \approx
|\Delta_\phi|/(1+\gamma)$, is controlled by the parameter $\gamma =
(2R_{\it NS}/R_N)(d/\xi_0)^2$, which could be small in short
junctions, $d \ll \xi_0$, even at large interface resistance, $R_{\it
NS} \gg R_{N}$. In contrast to the case of transmissive interfaces,
the supercurrent spectral density $I_s(E)$ is negative above the bulk
gap value.

In three-terminal interferometers, the supercurrent generally
coexists with the dissipative current flowing out from the injection
electrode. In such situation, we defined the non-equilibrium
Josephson current $j_s$ as the non-dissipative component of the
current flowing between the superconducting electrodes. In symmetric
junctions, within the subgap energy region, this component coincides
with its intuitive representation through the supercurrent spectral
density $I_s$ and the quasiparticle population imposed by
non-equilibrium injection, because the Andreev reflection blocks
quasiparticle exchange with equilibrium superconducting reservoirs.
In junctions with transmissive interfaces the Josephson current
becomes completely blocked at $eV = \Delta$ at zero temperature,
while in junctions with high-resistive interfaces, the Josephson
current undergoes inversion at $eV \lesssim \Delta$, which spreads
over a finite voltage interval. At the energies above the bulk energy
gap, $E > \Delta$, the population in the junction arms is basically
determined by the equilibrium population in the superconducting
reservoirs. By this reason, the Josephson current becomes voltage
independent at $eV > \Delta$.

We notice that spectroscopy of the supercurrent spectral density at
the subgap energies is possible at zero temperature, similar to the
tunnel spectroscopy of $N(E)$, because the derivative of the
Josephson current over applied voltage, $dj_s/dV$, is proportional to
$I_s(eV)$.

The critical current $j_c$ of the three-terminal junction, defined as
the maximum value of the transport current $j_T$ flowing through the
junction without creating a voltage drop, does not coincide with the
maximum in $j_s(\phi)$. This is due to the presence of phase
dependent injection current $j^V(\phi)$ which contributes to $j_c$,
along with the Josephson current, and adjusts its magnitude providing
zero voltage drop across the junction. At large voltage, where the
Josephson current is suppressed, the domain of the Josephson-like
regime is determined by the amplitude of phase oscillations of the
excess current in the injection electrode.

The behavior of the injection current is highly sensitive to the
quasiparticle spectrum of the junction and can be used to detect the
position of the phase-dependent proximity gap. In particular, the
differential conductance of the junction with perfect interfaces
exhibits sharp peaks at the bulk gap value, $eV = \Delta$, and at the
proximity gap, $eV = |\Delta_\phi|$; the latter becomes more
pronounced as the resistance of the injection lead increases.
Furthermore, the differential resistance exhibits full-scale
oscillations with the phase difference; at $eV < \Delta$, the shape
of the oscillations becomes rather complex, due to the interplay
between the position and amplitude of the proximity gap resonance.

\acknowledgments{ 

Support from VR and KVA (Sweden) is gratefully acknowledged.}


\begin{thebibliography}{99}

\bibitem{Lambert}
C.~J.~Lambert and R.~Raimondi, J. Phys. Condens. Matter. {\bf 10},
901, (1998).

\bibitem{Sam}
P.~Samuelsson, J.~Lantz, V.~S.~Shumeiko and G.~Wendin, Phys. Rev. B
{\bf 62}, 1309 (2000).

\bibitem{Hekking}
F.~W.~J.~Hekking and Yu.~V.~Nazarov, Phys. Rev. Lett. {\bf 71}, 1625
(1993); H.~Nakano and H.~Takayanagi, Phys. Rev. B {\bf 47}, 7986
(1993).

\bibitem{Wees}
B.~J.~van Wees, K.~M.~H.~Lenssen, and C.~J.~P.~M.~Harmans, Phys. Rev.
B {\bf 44}, 470 (1991).

\bibitem{Wendin}
G.~Wendin, and V.~S.~Shumeiko, Phys. Rev. B {\bf 53}, R6006 (1996).

\bibitem{VP} A.~F.~Volkov, and V.~V.~Pavlovski\u{\i}, JETP
Lett. {\bf 64}, 670 (1996).

\bibitem{VT} A.~F.~Volkov, and H.~Takayanagi, Phys. Rev. Lett. {\bf
76}, 4026 (1996); Phys. Rev. B {\bf 56}, 11184 (1997);
R.~Shaikhaidarov, A.~F.~Volkov, H.~Takayanagi, V.~T.~Petrashov, and
P.~Delsing,  {\it ibid.} {\bf 62}, R14649 (2000).

\bibitem{Petrashov}
V.~T.~Petrashov, V.~N.~Antonov, P.~Delsing and T.~Claeson,  Phys.
Rev. Lett. {\bf 70}, 347 (1993); P.~G.~N.~de Vegvar, T.~A.~Fulton,
W.~H.~Malinson, and R.~E.~Miller, {\it ibid.} {\bf 73}, 1416 (1994);
H.~Pothier, S.~Gueron, D.~Esteve, and M.~H.~Devoret, {\it ibid.} {\bf
73}, 2488 (1994).

\bibitem{VZ} A.~F.~Volkov, and A.~V.~Zaitsev, Phys. Rev. B {\bf 53},
9267 (1996); A.~V.~Zaitsev, A.~F.~Volkov, S.~W.~D.~Bailey, and
C.~J.~Lambert, {\it ibid.} {\bf 60}, 3559 (1999).

\bibitem{Morpurgo98}
A.~F.~Morpurgo, T.~M.~Klapwijk, and B.~J.~van Wees, Appl. Phys. Lett.
{\bf 72}, 966 (1998); J.~J.~A.~Baselmans, A.~F.~Morpurgo, B.~J.~van
Wees, and T.~M.~Klapwijk, Nature {\bf 397}, 43 (1999).

\bibitem{Baselmans02} J.~J.~A.~Baselmans, T.~T.~Heikkil\"a, B.~J.~van
Wees, and T.~M.~Klapwijk, Phys. Rev. Lett. {\bf 89}, 207002 (2002).

\bibitem{Huang} J.~Huang,
F.~Pierre, T.~T.~Heikkil\"a, F.~K.~Wilhelm, and N.~O.~Birge, Phys.
Rev. B {\bf 66}, 020507(R) (2002).

\bibitem{SamRap} P.~Samuelsson, V.~S.~Shumeiko, and G.~Wendin,
Phys. Rev. B {\bf 56}, R5763 (1997).

\bibitem{Bagwell}
L.~F.~Chang and P.~F.~Bagwell, Phys. Rev. B {\bf 55}, 12678 (1997).

\bibitem{SamPhysC}
P.~Samuelsson, \AA.~Ingerman, V.~Shumeiko, and  G.~Wendin, Physica C,
{\bf 352}, 1-4, 82 (2001).

\bibitem{Andreev} A.~F.~Andreev, Sov. Phys. JETP {\bf 19}, 1228
(1964).

\bibitem{VolkovSINIS} A.~F.~Volkov, Phys. Rev. Lett. {\bf 74}, 4730
(1995).

\bibitem{Wilhelm98} F.~K.~Wilhelm, G.~Sch\"{o}n, and A.~D.~Zaikin,
Phys. Rev. Lett. {\bf 81}, 1682 (1998).

\bibitem{Yip}
S.~K.~Yip, Phys. Rev. B {\bf 58}, 5803 (1998).

\bibitem{Heik} T.~T.~Heikkil\"a, J.~S\"arkk\"a, and F.~K.~Wilhelm,
Phys. Rev. B {\bf 66}, 184513 (2002).

\bibitem{VK} A.~F.~Volkov, and T.~M.~Klapwijk, Phys. Lett. A {\bf
168}, 217 (1992); A.~F.~Volkov, A.~V.~Zaitsev, and T.~M.~Klapwijk,
Physica C {\bf 210}, 21 (1993).

\bibitem{VolkovI} R.~Seviour, and A.~F.~Volkov, Phys. Rev. B {\bf 61},
9273 (2000).

\bibitem{Fur} A.~Furusaki, and M.~Tsukada, Phys. Rev. B {\bf 43},
10164 (1991).

\bibitem{nano} A.~Yu.~Kazumov, I.~I.~Khodos, P.~M.~Aiayan, and
C.~Col\-li\-ex, Euro\-phys. Lett. {\bf 34}, 429 (1996); A.~Bezryadin,
C.~N. Lau, and M.~Tinkham, Nature (London) {\bf 404}, 971 (2000).

\bibitem{LO} A.~I.~Larkin, and Yu.~N.~Ovchinnikov, Sov. Phys. JETP
{\bf 41}, 960 (1975); {\bf 46}, 155 (1977).

\bibitem{Naz} Yu.~V.~Nazarov, Phys. Rev. Lett. {\bf 73}, 1420
(1994); Superlattices Microstruct. {\bf 25}, 1221 (1999).

\bibitem{Usadel} K.~D.~Usadel, Phys. Rev. Lett. {\bf 25}, 507 (1970).

\bibitem{Mats} Similar solution for a diffusive point contact between
superconductors was found in Ref.~\onlinecite{KO} by Matsubara
technique.

\bibitem{note1} To next order, one should take into account the electron-hole
dephasing at finite energy [second term in the r.h.s. of
Eq.~(\ref{FirstInt})], which results in a small contribution to the
supercurrent from the continuum states, $E>\Delta$, where the
spectral density $I_s(E)$ is negative.\cite{Yip,Heik}

\bibitem{KL} M.~Yu.~Kupriyanov, and V.~F.~Lukichev, Sov.
Phys. JETP {\bf 67}, 89 (1988).

\bibitem{BBG} E.~V.~Bezuglyi, E.~N.~Bratus', and V.~P.~Galaiko,
Low Temp. Phys. {\bf 25}, 167 (1999).

\bibitem{Been} C.~W.~J.~Beenakker, Phys. Rev. Lett. {\bf 67}, 3836
(1991); {\em ibid.} {\bf 68}, 1442 (1992).

\bibitem{deJong} M.~J.~M.~de Jong, Phys. Rev. B {\bf 54}, 8144 (1996).

\bibitem{AB} V.~Ambegaokar, and A.~Baratoff, Phys. Rev. Lett. {\bf 10},
486 (1963).

\bibitem{note2} Similar formula has been applied for calculation of
the Josephson current in Ref.~\onlinecite{Huang}.

\bibitem{KO} I.~O.~Kulik, A.~N.~Omelyanchouk, Sov. J. Low Temp. Phys. {\bf
5}, 296 (1978).

\bibitem{AVZ} S.~N.~Artemenko, A.~F.~Volkov, and A.~V.~Zaitsev, Solid
State Commun. {\bf 30}, 771 (1979).

\bibitem{Reentrance} Yu.~N.~Nazarov, and T.~H.~Stoof, Phys. Rev. Lett.
{\bf 76}, 823 (1996); A.~F.~Volkov, N.~Allsopp, and C.~J.~Lambert, J.
Phys. Condens. Matter {\bf 8}, 45 (1996); A.~A.~Golubov,
F.~K.~Wilhelm, and A.~D.~Zaikin, Phys. Rev. B {\bf 55}, 1123 (1997).

\end{thebibliography}
\end{document}